\let \oldvec \vec
\documentclass[twocolumn]{IEEEtran}
\let \vec \oldvec
\usepackage{graphicx,amsmath,multirow}
\usepackage{algorithm}
\usepackage{algcompatible}
\usepackage{lmodern}
\usepackage{url}
\usepackage{color}

\renewcommand{\algorithmiccomment}[1]{\bgroup\hfill-~#1\egroup}

\begin{document}
\title{SSIDS: Semi-Supervised Intrusion Detection System by Extending the Logical Analysis of Data} 

%%%%---------------
\author{Tanmoy Kanti Das,  
\thanks{
T. K. Das is with
Department of Computer Applications,
National Institute of Technology Raipur,
INDIA.
Email:tkdas.mca@nitrr.ac.in}
Sugata Gangopadhyay,
\thanks{
S. Gangopadhyay is with
Department of CSE,
Indian Institute of Technology Roorkee, INDIA 247667,
Email:sugatfma@iitr.ac.in
}
Jianying Zhou
\thanks{
J.~Zhou is with
iTrust, Center for Research in Cyber Security,
Singapore University of Technology and Design, Singapore.
Email: jianying\_zhou@sutd.edu.sg}
}
%\date{}
%\journalname{}
\maketitle

\begin{abstract}
Prevention of cyber attacks on the critical network resources has become an important issue as the traditional Intrusion Detection Systems (IDSs) are no longer effective due to the high volume of network traffic and the deceptive patterns of network usage employed by the attackers. Lack of sufficient amount of labeled observations for the training of IDSs makes the semi-supervised IDSs a preferred choice. We propose a 
semi-supervised IDS by extending a data analysis technique known as Logical Analysis of Data, or LAD in short, which was proposed as a supervised learning approach. LAD uses partially defined Boolean functions (pdBf) and their extensions to find the positive and the negative patterns from the past observations for classification of future observations. We extend the LAD to make it semi-supervised to design an IDS. The proposed SSIDS consists of two phases: offline and online. The offline phase builds the classifier by identifying the behavior patterns of normal and abnormal network usage. Later, these patterns are transformed into rules for classification and the rules are used during the online phase for the detection of abnormal network behaviors. The performance of the proposed SSIDS is far better than the existing semi-supervised IDSs and comparable with the supervised IDSs as evident from the experimental results. 
\end{abstract}

 
\section{Introduction}
The Internet has evolved into a platform to deliver services from a platform to disseminate information. Consequently, misuse and policy violations by the attackers are routine affairs nowadays. Denning~\cite{den87} introduced the concept of detecting the cyber threats by constant monitoring of network audit trails using 
Intrusion Detection System (IDS) to discover abnormal patterns or signatures of network or system usage. Recent advancements in IDS are related to the use of machine learning and soft computing techniques that have reduced the high false positive rates which were observed in the earlier generations of IDSs~\cite{kim14}~\cite{tsai09}. The statistical models in the data mining techniques provide excellent intrusion detection capability to the designers of the existing IDSs which have increased their popularity. However, the inherent complications of IDSs such as competence, accuracy, and usability parameters make them unsuitable for deployment in a live system having high traffic volume. Further, the learning process of IDSs requires a large amount of training data which may not be always available, and it also requires a lot of computing power and time. Studies have revealed that it is difficult to handle high-speed network traffic by the existing IDSs due to their complex decision-making process. Attackers can take advantage of this shortcoming to hide their exploits and can overload an IDS using extraneous information while they are executing an attack. Therefore, building an efficient intrusion detection is vital for the security of the network system to prevent an attack in the shortest possible time.

A traditional IDS may discover network threats by matching current network behavior patterns with that of known attacks. The underlying assumption is that the behavior pattern in each attack is inherently different compared to the normal activity. Thus, only with the knowledge of normal behavior patterns, it may be possible to detect a new attack. However, the automatic generation of these patterns (or rules) is a challenging task, and most of the existing techniques require human intervention during pattern generation. Moreover, the lack of exhaustive prior knowledge (or labeled data) regarding the attacks makes this problem more challenging. It is advantageous for any IDS to consider unlabeled examples along with the available (may be small in number) labeled examples of the target class. This strategy helps in improving 
the accuracy of the IDSs against the new attacks. 
An IDS which can use both labeled and unlabeled examples is known as a semi-supervised IDS. Another important aspect of any intrusion detection system is the time required to detect abnormal activity. Detection in real time or near real time is preferred as it can prevent substantial damage to the resources. Thus, the primary objective of this work is to develop a {\it semi-supervised intrusion detection system} for near real-time detection of cyber threats.      

Numerous security breaches of computer networks have encouraged researchers and practitioners to design several Intrusion Detection Systems. For a comprehensive review, we refer to \cite{lia13}. Researchers have adopted various approaches to design IDSs, and a majority
of them modeled the design problem as a classification problem. In \cite{amb16}, a feature selection method is used with a standard classifier like SVM as the conventional classifiers perform poorly due to the presence of redundant or irrelevant features. Authors 
of \cite{wan17} also adopted a similar approach. Most of these designs share one common disadvantage, i.e., they follow a supervised learning approach. Recently, a new semi-supervised IDS has been proposed in~\cite{rana17}, and it outperforms the existing semi-supervised IDSs, but it suffers from the low accuracy of detection.   

It is essential to understand the behavior patterns of the known attacks, as well as the behaviors of normal activity to discover and prevent the attacks. Generation of patterns or signatures to model the normal as well as the abnormal activities is a tedious process, and it can be automatized using the application of LAD. Peter L. Hammer introduced the concept of logical analysis of data (or LAD) in the 
year $1986$~\cite{ham86} and subsequently developed it as a technique to find the useful rules and patterns from the past observations to classify new observations~\cite{bor00,cra88}. Patterns (or rules) can provide a very efficient way to solve various problems in different application areas, e.g., classification, development of rule-based decision support system, feature selection, medical diagnosis, network traffic analysis, etc. The initial versions of LAD~\cite{ale07,cra88,ham86} were designed to work with the binary data having either of the two labels, i.e., positive or negative. Thus, the data or observations were part of a two-class system. A specific goal of LAD is to learn the logical patterns which set apart observations of a class from the rest of the classes. 

LAD has been used to analyze problems involving medical data. 
A typical dataset consists of two disjoint sets $\Omega^+, \Omega^-$ which represent a set of observations consisting of positive and negative examples, respectively. Here, each observation is a vector consisting of different attribute values. In the domain of medical data analysis, each vector represents the medical record of a patient, and the patients in $\Omega^+$ have a specific medical condition. On the other hand, $\Omega^-$ represents the medical records of the patients who do not have that condition. Subsequently, if a new vector / patient is given, one has to decide whether the new vector belongs to  $\Omega^+$ or $\Omega^-$, i.e., one has to determine whether the patient has the particular medical condition or not. Thus, in this example, the medical diagnosis problem can be interpreted as a two-class classification problem. The central theme of LAD is the selection of such patterns (or rules) which can collectively classify all the known observations. LAD stands out in comparison with other classification methods since a pattern can explain the classification outcome to human experts using formal reasoning.

Conventional LAD requires labeled examples for the pattern or rule generation. However, there exist several application domains (e.g., intrusion detection system, fraud detection, document clustering, etc.) where the existence of labeled examples are rare or insufficient. To harness the strength of LAD in these application domains, one needs to extend LAD for unsupervised and semi-supervised pattern generation~\cite{bruni15}. Here, we introduce a preprocessing methodology using which we can extend the LAD in such a manner that it can use unlabeled observations along with the labeled observations for pattern generation. Consequently, it acts like a {\it semi-supervised} learning approach. The central theme is to use the classical LAD to generate initial positive and negative patterns from the available labeled observations. Once the patterns are available, we measure the closeness of the unlabeled observations with the initial positive or negative patterns using balance score. The observations with high positive balance score are labeled as the positive observations and the observations having high negative balance score are labeled as the negative examples. Once labels are generated, the standard LAD can be used as it is. We have used this approach successfully in the design of a new {\it semi-supervised} and {\it lightweight} Intrusion Detection System (IDS) which outperforms the existing methods in terms of accuracy and requirement of computational power.  

Creation of signatures or patterns to model the normal as well as the abnormal network activities can be accomplished using the semi-supervised LAD (or S-LAD in short), and in this effort, we have used  S-LAD to design a semi-supervised IDS. Here,  S-LAD is used to generate the patterns which can differentiate the normal activities from the malicious activities, and these patterns are later converted to rules for the classification of unknown network behavior(s). The proposed SSIDS has two phases, the offline phase is used to design a rule-based classifier. This phase uses historical observations, both labeled and unlabeled, to find the patterns or rules of classification, and require a significant amount of processing power. Once the classification rules are generated, the online phase uses those rules to classify any new observation. The online  phase requires much less processing power than the offline phase, and it can detect threats in near real-time. The accuracy of proposed semi-supervised IDS is much better than any state-of-the-art semi-supervised IDS and comparable with the supervised IDSs. 

The main contributions of the proposed paper are: (1) a new implementation of LAD having extensively modified pattern generation algorithm; (2) a new strategy to extend LAD that is suitable for the design of semi-supervised classifiers; (3) a LAD-based design of a lightweight semi-supervised intrusion detection system that outperforms any existing semi-supervised IDSs.      

The rest of the paper is organized as follows. Next section gives a brief description of our modified implementation of LAD and Section~\ref{slad} describes the proposed method to extend  LAD to the semi-supervised LAD. Details of the proposed SSIDS is available in the Section~\ref{sids}. Performance evaluation and comparative results are available in the Section~\ref{expr} and we conclude the paper in the Section~\ref{sec-con}.

\section{Proposed Implementation of LAD}
\label{lad}
LAD is a data analysis technique which is inspired by the combinatorial optimization methods. 
%It has been used in numerous applications, e.g., economics and business, oil exploration, medical diagnosis etc. 
As pointed out earlier, the initial version of LAD was designed to work with the binary data only. Let us first briefly describe the basic steps of LAD when it is applied to the binary data. An observation having $n$ attributes may be represented as a binary vector of length $n+1$ as the last bit (a.k.a. the class label) indicates whether it is a member of $\Omega^+$ or $\Omega^-$. Thus, the set of binary observations $\Omega$ ($=\Omega^+ \cup \Omega^- \subseteq \{0,1\}^n$) can be represented by a partially defined Boolean function (pdBf in short) $\phi$, indicating a mapping of $\Omega \rightarrow \{0,1\}$. The goal of LAD is to find an extension $f$ of the pdBf $\phi$ which can classify all the unknown vectors in the sample space. 
However, this goal is clearly unachievable and we try to find an approximate extension $f^\prime$ of $f$. $f^\prime$ should approximate $f$ as closely as possible based on the several optimality criteria. Normally, the extension is represented in a disjunctive normal form (DNF). In brief, the LAD involves following steps~\cite{ale07}.

{\small 
\begin{enumerate}
    \item {\it Binarization of Observations. We have used a slightly modified implementation of binarization here.}
	\item {\it Elimination of Redundancy (or Support Sets Generation).}
	\item {\it Pattern Generation. Our extensively modified pattern generation algorithm makes the 'Theory Formation' step redundant.}
	\item {\it Theory Formation. We have omitted this step.}% in our implementation.} 
	\item {\it Classifier Design and Validation}.
\end{enumerate}
}
There are many application domains from the finance to the medical where the naturally occurring data are not binary~\cite{ale07,bor97}. Thus, to apply LAD in those domains, a method to convert any data to binary is discussed in the subsection~\ref{binr}. Moreover, we have modified the original pattern generation algorithm in such a manner that the coverages of every pair of patterns have a very low intersection. Thus, the step ``theory formation" is 
no longer required. Recently, a technique to produce internally orthogonal patterns (i.e., the coverages of every pair of patterns have empty intersection) is also reported in~\cite{burni18}. 

\subsection{Binarization of Observations}
\label{binr}
A threshold (a.k.a. cut-point) based method was proposed to convert the numerical data to binary. % which preserves the inherent characteristics of the data. 
Any numerical attribute $x$ is associated with two types
of Boolean variables, i.e. the {\it level variables} and the {\it interval variables}. %For details we refer to \cite{bor00}. 
Level variables are related to the cut-points and indicate whether the original attribute value is greater than or less than the given cut-point $\beta$. For each cut-point $\beta$, we create a Boolean variable $b(x, \beta)$ such that

{\small
\begin{equation}
  b(x, \beta)=\begin{cases}
    1, & \text{if $x \ge \beta$}.\\
    0, & \text{otherwise}.
  \end{cases}
\end{equation}
}
\noindent Similarly, interval variables are created for each pair of cut-points $\beta_1$ and $\beta_2$ and represented by Boolean
variable $b(x, \beta_1, \beta_2)$ such that
{\small
\begin{equation}
  b(x, \beta_1, \beta_2)=\begin{cases}
    1, & \text{if $ \beta_1 \le x < \beta_2$}.\\
    0, & \text{otherwise}.
  \end{cases}
\end{equation}
}
We are yet to discuss how the cut-points are determined. The cut-points should be chosen carefully such that the resultant pdBf should have an extension in the class of all Boolean functions $\mathcal{C}_{ALL}$~\cite{bor97}. Let us consider the numerical attribute $x$ having $k+1$ distinct values present in the observations and the attribute values are ordered such that $x_0>x_1>\ldots >x_k$. We introduce a cut-point between 
$x_i$ and $x_{i+1}$ if they belong to different classes. The resulting pdBf is referred to as the {\it master} pdBf if we create cut-point for each pair of values. 
%We may have to introduce $k$ many cut-points if every pair of consecutive observations $x_i$ and $x_{i+1}$ are the members of different classes and it is the largest possible set of cut-points for the feature $x$. %If we have to introduce the largest possible set of cut-points for every attribute, then the resultant pdBf is known as the {\it master} pdBf. 
%The cut-points are chosen carefully in a manner that one can distinguish between the positive and the negative observations. 
%
Note that, the resultant master pdBf 
has extension in $\mathcal{C}_{ALL}$ if and only if $\Omega^+ \cap \Omega^- = \emptyset$. 

The process for selection of cut-points is explained below using an example from~\cite{tkd19}.
%However, the employed cut-point selection process is different from what has been proposed in~\cite{bor97,bor00}. 
The original dataset presented in the Table~\ref{dset1} is converted to the Table~\ref{dset2} by adding the class labels (or truth values of pdBf). Those observations that are the members of $\Omega^+$ have $1$ as the class label and rest of the observations have $0$ as the class labels. Now, if we want to convert the numeric attribute $A$ to binary, we form another dataset as represented in the Table~\ref{dset3}. Next, we sort the dataset over the attribute $A$ to get a new dataset $D$ that is presented in the Table~\ref{dset4}. After that, we apply the following steps to get the cut points.

\begin{enumerate}
	\item Preprocessing of $D$: This step is a slight modification of the 
	usual technique used in \cite{bor97,bor00}, and other related papers. 
	If two or more consecutive observations have the same attribute value $v_i$ but different class labels, 
	remove all those observations except one observation. Now, we change the existing class label of  $v_i$	to a new and unique class label which does not appear in $D$ and include that  in the set of class labels of $D$. Refer to Table~\ref{dset5}. 
		%\item Repeat the last step until only unique attribute values are left.
	\item Now, if two consecutive observations $A_i$ and $A_{i+1}$ have different class labels, introduce a new 
	cut-point $\beta^A_j$ as
	$$\beta^A_j = \frac{1}{2} (A_i + A_{i+1})$$
\end{enumerate}

\noindent If we follow the above mentioned steps, the obtained cut-points are $\beta^A_1 = 3.05$, $\beta^A_2 = 2.45$, $\beta^A_3 = 1.65$. Thus, we will have six Boolean variables consisting of three level variables and three interval variables corresponding to these cut-points. 
%After the application of previous steps over all the attributes of the Table~\ref{dset2}, the binary dataset obtained is presented in the Table~\ref{bintab} which is an example of a partially defined Boolean function.
\noindent
		\begin{table*}[!ht]
		\resizebox{1.7\columnwidth}{!}{
		\small
		\begin{minipage}{0.32\textwidth}
		\resizebox{1.0\textwidth}{!}{
		\small
		\centering
		\begin{tabular}{|c|c|c|c|}
		  \hline
			Attributes&$A$&$B$&$C$\\
			&&&\\
			\hline
			$\Omega^+$:positive&3.5&3.8&2.8\\
			%examples & & &  \\
			\cline{2-4}
			 examples & 2.6 & 1.6 & 5.2 \\
			\cline{2-4}
			 & 1.0 & 2.1 & 3.8 \\
			\hline
			\hline
			$\Omega^-$:negative  & 3.5 & 1.6 & 3.8 \\
			\cline{2-4}
			examples & 2.3 & 2.1 & 1.0 \\
			\hline
			\multicolumn{4}{c}{}
			\end{tabular}
			}
			\caption{}
			%\caption{A Numerical Dataset}
			\label{dset1}
			%\end{table}
		\end{minipage}
		%\hfill
		\begin{minipage}{0.26\textwidth}
		\resizebox{1.0\textwidth}{!}{
		%\small
		\centering
		\begin{tabular}{|c|c|c|c|}
		  \hline
		 $A$ & $B$ & $C$ & Class\\
		& & & Labels\\
			\hline
			 3.5 & 3.8 & 2.8 & 1 \\
			\hline
			 2.6 & 1.6 & 5.2 & 1 \\
		\hline
			 1.0 & 2.1 & 3.8 & 1\\
			\hline
			\hline
			 3.5 & 1.6 & 3.8 & 0 \\
		  \hline
			2.3 & 2.1 & 1.0 & 0\\
			\hline
			\multicolumn{4}{c}{}
			\end{tabular}
			}
			\caption{}
			\label{dset2}
			%\caption{The same numerical dataset of table~\ref{dset1} with class labels.} 
			%\end{table}
	\end{minipage}
	%\hfill
	\begin{minipage}{0.155\textwidth}
	\resizebox{\textwidth}{!}{
		%\small
		\centering
		\begin{tabular}{|c|c|}
		  \hline
		 $A$ & Class\\
		& Labels\\
			\hline
			 3.5  & 1 \\
			\hline
			 2.6  & 1 \\
		\hline
			 1.0  & 1\\
			\hline
			\hline
			 3.5  & 0 \\
		  \hline
			2.3  & 0\\
			\hline
			\multicolumn{2}{c}{}
			\end{tabular}
		  }
			\caption{}
			\label{dset3}
			%\caption{The same numerical dataset of table~\ref{dset1} with class labels.} 
			%\end{table}
	\end{minipage}
	%\hfill
	\begin{minipage}{0.16\textwidth}
	\resizebox{1.0\textwidth}{!}{
		%\small
		\centering
		\begin{tabular}{|c|c|}
		  \hline
		 $A$ & Class \\
		& Labels\\
			\hline
			 3.5  & 1 \\
			\hline
			 3.5  & 0 \\
			\hline
			 2.6  & 1 \\
		  %\hline
			\hline
			2.3  & 0\\
			\hline
			 1.0  & 1\\
			\hline
			\multicolumn{2}{c}{}
			\end{tabular}
			}
			\caption{}
			\label{dset4}
			%\caption{The same numerical dataset of table~\ref{dset1} with class labels.} 
			%\end{table}
	\end{minipage}
	\begin{minipage}{0.185\textwidth}
	\resizebox{\textwidth}{!}{
	%\small
	\centering
	\begin{tabular}{|c|c|}
		\hline
		$A$ & Class\\
		& Labels\\
		\hline
		3.5  & 2 \\
		\hline
		2.6  & 1 \\
		%\hline
		\hline
		2.3  & 0\\
		\hline
		1.0  & 1\\
		\hline
		\multicolumn{2}{c}{}
	\end{tabular}
	}
	\caption{}
	\label{dset5}
	%\caption{The same numerical dataset of table~\ref{dset1} with class labels.} 
	%\end{table}
\end{minipage}
}
\end{table*}		
A ``nominal" or descriptive attribute $x$ can be converted into binary very easily by relating each possible value $v_i$ of $x$ with a Boolean variable $b(x,v_i)$ such that
\begin{equation}
  b(x, v_i)=\begin{cases}
    1, & \text{if $ x = v_i$}.\\
    0, & \text{otherwise}.
  \end{cases}
\end{equation}
%Note that, this process may not be required if the number of possible values of the descriptive attribute is two and thus, it may be represented by a single Boolean variable. 

\subsection{Support sets generation}
Binary dataset obtained through the binarization or any other process may contain redundant attributes. A set $S$ of binary attributes is termed as a {\it support set} if the projections $\Omega^+_S$ and $\Omega^-_S$  of $\Omega^+$ and $\Omega^-$, respectively, are such that $\Omega^+_S \cap \Omega^-_S = \emptyset$. 
A support  set is termed {\it minimal} if elimination any of its constituent attributes leads to
 $\Omega^+_S \cap \Omega^-_S \ne \emptyset$.
 	Finding the minimal support set of a binary dataset, like Table~\ref{bintab} (see Appendix), is equivalent of solving a set covering problem. A detailed discussion on the support set, minimal support set and a few algorithms to solve the set covering problem can be found in ~\cite{alm94,cra88,ham86}. Here, we have used the {\it ``Mutual-Information-Greedy" algorithm} proposed in~\cite{alm94} to solve the set covering problem in our implementation. Note that, our implementation produces the set $S$ in a manner such that the constituent binary attributes are ordered according to their discriminating power and it helps us to achieve the simplicity objective which is mentioned in the description of LAD. Following binary feature variables are selected if we apply the said algorithm:  $S=\{b_{15},b_8,b_1,b_2\}$.    	
 %\comm{To be defined $T^{\prime}({\Omega}^+_S)$}

\subsection{Modified pattern generation method}
Let us first recall a few common Boolean terminologies that we may require to describe the pattern generation process.
A Boolean variable or its negation is known as {\it literals} and conjunction of such literals is called a {\it term}.
The number of literals present in a term $T$ is known as its {\it degree}. The {\it characteristic term} of a point
$p \in \{0,1\}^n$ is the unique term of degree $n$, such that $T(p)=1$. The term $T$ is said to {\it cover} the point 
$p$ if $T(p)=1$. A term $T$ is called a {\it positive pattern} of a given dataset $(\Omega_S^+, \Omega_S^-)$ if
\begin{enumerate}
\item $T(p)=0$ for every point $p \in \Omega_S^-$.
\item $T(p)=1$ for at least one point $p \in \Omega_S^+$. 
\end{enumerate}
Similarly, one can define the negative patterns. {Here, $T(\Omega_S)$ is defined as $T(\Omega_S)= \bigcup \limits_{p \in \Omega_S} T(p)$.} 
Both the positive and the negative patterns play a significant role in any
LAD based classifier. A {\it positive pattern} is defined as a subcube of the unit cube that intersects $\Omega^+_S$ but is disjoint from
$\Omega^-_S$. A {\it negative pattern} is defined as a subcube of 
the unit cube that intersects $\Omega^-_S$ but is disjoint from 
$\Omega^+_S$. Consequently, we have a symmetric pattern generation procedure.
% that can follow a ``top-down'' or ``bottom-up'' approach. 
%In the bottom-up approach, computation starts with the terms of degree one and if a
%term covers a positive point (or observation) but no negative point, then it is a pattern. However, if no pattern is found, literals
%are added to the terms as long as required. On the contrary, the top-down approach begins computation with the characteristics term of a point.
%Such terms are definitely a pattern and if we remove some literals from such terms, they may still remain a pattern. In such a scenario,
%the top-down approach methodically purges literals from the term to arrive at a minimal pattern also known as the {\it prime pattern}.  
In this paper, we have used an extensively modified and optimized version of the pattern generation technique that has been proposed by Boros et al.~\cite{bor00}.
 
{\small
	\center
	\begin{algorithm}[!ht]
		\centering
		\begin{algorithmic}[1]
			\STATEx{Input: \quad ${\Omega}_S^+$, ${\Omega}_S^- \subset \{0,1\}^n$ - Sets of positive and negative observations in binary.}
			\STATEx{$\hat{d}$ \hspace{1mm}- Maximum degree of generated patterns.}
			\STATEx{$k$ \hspace{1mm}- Minimum number of observations covered by a generated pattern.} 
			\STATEx{Output: \hspace{0mm} $\chi$ \hspace{1cm}- Set of prime patterns.}
			\STATE{$\chi=\emptyset$.}
			\STATE{$\mathcal{G}_0=\{\emptyset\}$.}
			\FOR{ $d=1,\ldots,\hat{d}$}
			\IF {$d<\hat{d}$}
			\STATE {$\mathcal{G}_d=\emptyset$.} \COMMENT {$\mathcal{G}_{\hat{d}}$ is not required.}
			\ENDIF
			\FOR {$\tau \in \mathcal{G}_{d-1}$}
			\STATE{ $p=$ maximum index of the literal in $\tau$.}
			\FOR{ $s=p+1, \ldots, n$}
			\FOR{$l_{new} \in \{l_s,\bar{l}_s\}$}
			\STATE{$\tau^{\prime} = \tau \Vert l_{new}$.}
			\FOR {$i=1$ to $d-1$}
			\STATE {$\tau^{\prime \prime} = $ remove $i$\textsuperscript{th} literal from $\tau^{\prime}$.}
			\IF {$\tau^{\prime \prime} \notin \mathcal{G}_{d-1}$}
			\STATE {go to Step~\ref{dia}.}
			\ENDIF
			
			\ENDFOR
			\IF {$k \leq \sum_{y \in {\Omega}^+_S}\tau^{\prime}(y)$} \COMMENT {$\tau^{\prime}$ covers at least $k$ many positive observations.}
			\label{msup}
			\IF {$1 \notin \tau^{\prime}({\Omega}^-_S)$} \COMMENT {$\tau^{\prime}$ covers no negative observation.}
			\STATE{$\chi=\chi \cup \{\tau^{\prime}\}$.}
			\STATE{ Remove the points (or observations) covered by $\tau^{\prime}$ from ${\Omega}^+_S$ .} \label{thfm}
			\ELSIF {$d<\hat{d}$}
			\STATE{$\mathcal{G}_d= \mathcal{G}_d \cup \{\tau^{\prime}\}$.}
			\ENDIF
			\ENDIF
			\ENDFOR \label{dia} 
			
			\ENDFOR
			
			\ENDFOR
			
			\ENDFOR
		\end{algorithmic}
		\caption{\small Positive prime pattern enumeration algorithm.}% based on~\cite{bor00}.}
		\label{algo-pat}
	\end{algorithm}
}

We have made two major changes in  Algorithm~\ref{algo-pat} for pattern generation  over the algorithm
proposed in~\cite{bor00}. Steps~\ref{msup} and~\ref{thfm} are different from the original algorithm and 
Step~\ref{thfm} increases the probability that a point or observation is only covered by a single pattern instead of multiple patterns. 
We expect that the majority of the observations will be covered by a unique pattern. 
Thus, we no longer require the `theory formation' step to select the most suitable pattern to cover an observation.  
In Step~\ref{msup}, we have ensured that a pattern is selected if and only if it covers 
at least $k$ many positive observations.
This ensures that a selected pattern occurs frequently in the dataset. 
One major drawback of this approach is that if $k>1$, then it may so happen  that all the observations present
in the dataset may not be covered by the selected set of patterns. However, a properly chosen value of $k$ 
ensures that more than $95\%$ of the observations are covered.
Note that, the negative prime patterns can also be generated in a similar fashion.
If we apply the algorithm~\ref{algo-pat} over the projection 
$S=\{ b_{15},b_8,b_1,b_2\}$ of the binary dataset presented in the Table~\ref{bintab} (see Appendix), following positive patterns are generated:
(i) $b_2 b_8$, (ii) $b_2 \bar{b}_1$, (iii) $\bar{b}_2 b_{15}$ using $k=1$ and the corresponding negative patterns 
are (i) $\bar{b}_2 \bar{b}_{15}$, (ii) $b_2 b_{15}$. 

\subsection{Design of Classifier}
\label{dclfr}
The patterns which are generated using Algorithm~\ref{algo-pat}, are transformed into rules and later these rules are
used to build a classifier. The rule generation process is trivial and it's explained using an example. Let us take the first positive
pattern $b_2 b_8$. The meaning of $b_2$ is whether $(A \ge 2.45)$ is true or false as evident from the Table~\ref{bintab}. Similarly, the meaning of 
$\bar{b}_2$ is whether $\neg(A \ge 2.45)$ is true or false. Consequently, the rule generated from the pattern $b_2 b_8$ is 
$(A \ge 2.45) \land  (B \ge 1.85$) $\implies$  $\mathcal L=1$. The corresponding pseudo-code is as follows. 
\vspace{1mm}
\noindent
\newline 
{\bf if} $(A \ge 2.45) \land  (B \ge 1.85)$~{\bf then} \newline 
 \hspace*{5mm}Class label $\mathcal L = 1$ \newline
{\bf end if} 
\vspace{1mm}
\noindent
\newline We can combine more than one positive rule into an {\it `if else-if else'} structure to design a classifier. 
Similarly, one can build a classifier using the negative patterns also. Hybrid classifiers can use both the positive and the 
negative rules to design a classifier. A simple classifier using the positive patterns is presented below.

{
\renewcommand{\thealgorithm}{}
\floatname{algorithm}{}
\begin{algorithm}[!ht]
\centering
\small
\begin{algorithmic}[1]
\STATEx{Input: Observation consisting of attribute $A,B,C$.}
\STATEx{Output: Class label $\mathcal L$.}
\IF {($A \ge 2.45) \land  (B \ge 1.85$)}
	\STATE{Class label $\mathcal L = 1$.}
\ELSIF {($A \ge 2.45) \land \neg(A \ge 3.05$)}
   \STATE{Class label $\mathcal L = 1$.}
\ELSIF {$\neg (A \ge 2.45) \land ( 3.3 \leq C < 4.5)$}
   \STATE{Class label $\mathcal L = 1$.}
\ELSE
       \STATE{Class label $\mathcal L = 0$.}
\ENDIF
\end{algorithmic}
\caption{Simple Classifier.}
\label{clsfr}
\end{algorithm}
} 

In general, a new observation $x$ is classified as {\it positive} if at least a positive pattern covers it and no negative pattern
covers it. Similar definition is possible for {\it negative} observations.
However, in the `Simple Classifier', we have relaxed this criterion
and we consider $x$ as negative if it is not covered by any positive patterns. Another classification strategy that has worked well
in our experiment is based on {\it balance score}\cite{ale07}. The balance score is the linear combination of positive ($P_i$) and
negative ($N_i$) patterns and defined as :
\begin{equation}
\Delta(x) = \frac{1}{q}\sum_{l=1}^{q} P_l(x) - \frac{1}{r}\sum_{i=1}^{r} N_i(x)
\label{diseq}
\end{equation}     
The classification $\eta(x)$ of the new observations $x$ is given by
\begin{equation}
  \eta(x)=\begin{cases}
    1, & \text{if $ \Delta(x) > 0$}.\\
    0, & \text{if $ \Delta(x) < 0$}.\\
		\epsilon, & \text{if $ \Delta(x) = 0$}. ~\text{Here, $\epsilon$ indicates unclassified.} \\
  \end{cases}
\label{clseq}
\end{equation}
%\noindent 

\section{Extension of LAD}
\label{slad}
Majority of the applications of the LAD which are available in the existing literature~\cite{ale07}, work with the labeled data during the 
classifier design phase. There
are many applications where a plethora of data are available which are unlabeled or partially labeled. These applications require
semi-supervised or unsupervised pattern generation approach. One such application is {\it intrusion detection system} where
the lightweight classification methods designed using the LAD are desirable. However, the dearth of labeled observations makes it difficult
for the development of a LAD based solution. In this effort, we propose a pre-processing method which can label the available
unlabeled data. However, the proposed method requires that some labeled data are available during the design of classifiers. Thus,
the method is akin to a semi-supervised learning approach~\cite{zhu05,zhu09}.

The process of class label generation is very simple and it uses a standard LAD based classifier~\cite{bor00} having 
{\it balance score}~\cite{ale07} as a discriminant function to classify an unlabeled observation. First, we design a 
balance score based classifier using the set of available labeled observations D\textsubscript{L}. 
Later, we classify each observation in the unlabeled dataset using the balance score based classifier. However, we replace the 
classifier described in the Equation~\ref{clseq} by the Equation~\ref{clseq2}. Thus, we keep those observations unlabeled which are having
very low balance score and those observations are also omitted form farther processing. Basically, we are ensuring that if a given observation has a strong affinity towards the positive or negative patterns, then only the observation
is classified/labeled during the labeling process. 
\begin{equation}
  \eta^\prime(x)=\begin{cases}
    1, & \text{if $ \Delta(x) > \tau_1$}.\\
    0, & \text{if $ \Delta(x) < \tau_0$}.\\
		\epsilon, & \text{if $ \tau_0 \leq \Delta(x) \leq \tau_1$}.\\
  \end{cases}
\label{clseq2}
\end{equation}

We have evaluated the performance of the said strategy using the KDDTrain\_20 percent dataset which is part of the NSL-KDD dataset. The KDDTrain\_20percent dataset
consists of $25,192$ observations and we have partitioned the dataset into two parts. The first part D\textsubscript{L} consists of $5000$ randomly selected observations, and the second part D\textsubscript{UL} consists of the rest of the observations. We have removed the labels from the observations of D\textsubscript{UL}. Afterward, D\textsubscript{L} is used to design a classifier based on the Equation~\ref{clseq2}. This classifier latter used for the classification of 
D\textsubscript{UL} and the output of the labeling process is a dataset D\textsubscript{L}\textsuperscript{$\prime$} which consists of all the labeled examples from the D\textsubscript{UL}. 
The results are summarized in the Table~\ref{tab-kmns}. It is obvious that
any error in the labeling process will have a cascading effect on the performance of Algorithm~\ref{algo-ids}. On the
other hand, the unlabeled samples (marked as $\epsilon$) would have no such consequence on the performance of the proposed SSIDS.
Thus, while reporting the accuracy of the labeling process, we have considered the labeled samples only.  
It is clear from the Table~\ref{tab-kmns} that the number of observations that are currently labeled is $17601 + 5000 = 22601$ and these many observations would be used for farther processing. One important aspect
that remains to be discussed is the values of $\tau_0$, and $\tau_1$. We have used {\boldmath $\tau_0 =-0.021$} and 
{\boldmath$\tau_1 = 0.24$} in our experiments. 
We have arrived at these values after analyzing the outcome of the labeling process on the training dataset D\textsubscript{L}.  
\begin{table}[ht]
\center
\begin{tabular}{|c|c|c|c|c|}
\hline
%\#$D_L$ & \#$D_{UL}$ & \#Labeled & \# Wrongly & \#Unlabeled\\
\#D\textsubscript{UL} & \multicolumn{3}{|c|}{Labeled} & \#Unlabeled$(\epsilon)$\\
\cline{2-4}
 & \# Correctly & \#Wrongly & Accuracy &\\
\hline
20192 & 17333 & 268 &98.48\%&2591 \\
\hline
\multicolumn{5}{c}{}
\end{tabular}
\caption{Results related to the labeling of D\textsubscript{UL}.}
\label{tab-kmns}
\end{table}
%\vspace*{-3mm}
Following the introduction of this pre-processing step, the steps of a semi-supervised LAD (or {\bf S-LAD}) are as follows.

{\small
\begin{enumerate}
  \item {Class label (or truth value) generation.}
	\item {Binarization.}
	\item {Elimination of redundancy (or Support sets generation).}	
	\item {Pattern generation.}
	%\item {Theory formation.} 
	\item {Classifier design and validation.}
\end{enumerate}
}
\section{Design of a Semi-Supervised IDS using S-LAD}
\label{sids}
Organizations and governments are increasingly using the Internet to deliver services, and the attackers are trying to
gain unfair advantages from it by misusing the network resources. Denning~\cite{den87} introduced the concept
of detecting the cyber threats by constant monitoring of the network audit trails using the intrusion detection systems.
The intrusion can be defined as the set of actions that seek to undermine the availability, integrity or confidentiality
of a network resource~\cite{den87,her09,yan15}. Traditional IDSs that are used to minimize such risks can be categorized into two: (i) {\it anomaly based}, (ii) {\it misuse based} (a.k.a. signature based). The anomaly based IDSs build a model of normal activity, and any deviation from the model is considered as an intrusion. On the contrary, misuse based models generate signatures from the past attacks to analyze existing network activity. It was observed that the misuse based models are vulnerable to ``zero day'' attacks~\cite{mukk05}. Our proposed technique is unique in the sense that it can be used as either a misuse based or an anomaly based model. Hybridization is also possible in our proposed technique.

\subsection{Proposed Intrusion Detection System}
\begin{figure}[ht]
\center
\includegraphics[width=8cm]{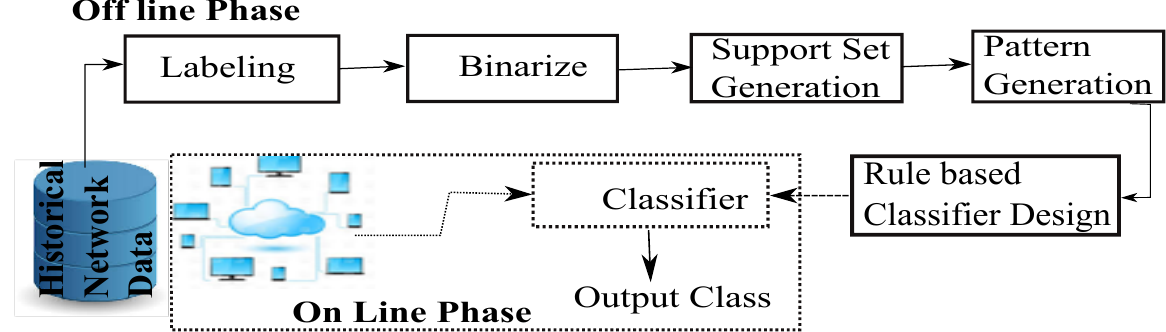}
\caption{Block Diagram of the proposed SSIDS}
\label{fig-ids}
\end{figure}

The proposed SSIDS is presented in the Figure~\ref{fig-ids}. It consists of two major phases, i.e., the offline phase and the online phase. 
The offline phase uses an S-LAD to design a classifier which online phase uses for real-time detection of any abnormal activity using the data that describe the network traffic. It is obvious that the offline phase should run at least once before the online phase is used
to detect any abnormal activity. The offline phase may be set up to run at a regular interval of time to upgrade the classifier with the new patterns or rules. Let us now summarize the steps of the offline phase in Algorithm~\ref{algo-ids}. Note that, the
Step~\ref{bls} of Algorithm~\ref{algo-ids} implicitly uses the Steps~\ref{sbin} to \ref{spgen} to build the classifier.
The online phase is very simple as it uses the classifier generated in the offline phase for the classification of new observations. 
\renewcommand{\thealgorithm}{2}
\begin{algorithm}[!ht]
\centering
\small
\begin{algorithmic}[1]
		\STATEx {{Input}: Historical dataset consisting of labeled and unlabeled data.}
		\STATEx {{Output}: Rule based classifier for the online phase.}
		\STATE {Read the historical dataset D\textsubscript{L} and D\textsubscript{UL}.}
		\STATE \label{bls}{Using D\textsubscript{L}, build a standard LAD Classifier based on the balance score (i.e., Equation~\ref{clseq2}).}
		\STATE {Using the classifier from the previous step, label the dataset D\textsubscript{UL} 
		to generate D\textsubscript{L}\textsuperscript{$\prime$} and  
		$\Omega=$D\textsubscript{L} $\cup$ D\textsubscript{L}\textsuperscript{$\prime$}.}
		\STATE \label{sbin}{Binarize $\Omega$ using the process described in Subsection~\ref{binr}.}
		\STATE {Generate support set $S$ from the binary dataset.}
		\STATE \label{spgen}{Generate positive and negative patterns (i.e., rules) using Algorithm~\ref{algo-pat}.}
		\STATE \label{ana}{Design a classifier from the generated patterns following the example of 'Simple Classifier' from Subsection~\ref{dclfr}.}
\end{algorithmic}
\caption{Steps of Offline Phase of IDS}
\label{algo-ids}
\end{algorithm}
One can use the positive rules only to build a classifier in Step~\ref{ana} of Algorithm~\ref{algo-ids},
then the IDS can be termed as anomaly-based. 
On the other hand, if it uses only the negative rules, the design is similar to a signature-based IDS. 

%\vspace{-mm} 

\section{Performance Evaluations}
\label{expr}
Most widely used datasets for validation of IDSs are NSL-KDD~\cite{nslkdd09} and KDDCUP'99~\cite{kddcup99}. NSL-KDD is a modified version of the KDDCUP'99 dataset and we have used the NSL-KDD dataset in all our experiments. Both the datasets  consist of $41$ features along with a class label for each observation. These features are categorized into four different classes and they are (i) basic features, (ii) content features, (iii) time-based traffic features, (iv) host-based traffic features. Here, the {\it basic features} are extracted from the TCP/IP connections without scanning the packets and there are nine such features in the NSL-KDD dataset. On the other hand, features which are extracted after inspecting the payloads of a TCP/IP connection are known as the {\it content features} and there are $13$ such features present in the dataset. A detailed description of the features is available in  the Table~\ref{tab-inpf}. There are different types of attacks present in the dataset but we have clubbed them to one and consider them as ``attack'' only. Thus, there are two types of class labels that
we have considered in our experiments and they are ``normal'' and ``attack''. We have used the KDDTrain\_20percent dataset which is a part of the NSL-KDD dataset to build the classifier in the offline phase. The KDDTest\textsuperscript{+} and the KDDTest\textsuperscript{-21} 
have been used during the online phase for validation testing. 
The details of the experimental setup are presented in Subsection~\ref{exp-setup}.
\begin{table}[ht]
\center
\resizebox{1\columnwidth}{!}{
\small \bf
\begin{tabular}{|l|c|l|l||l|c|l|l|}
\hline
\rotatebox{90}{Feature~~}\rotatebox{90}{Type} &\rotatebox{90}{Col. No}& Input Feature &\rotatebox{90}{Data Type} & \rotatebox{90}{Feature~~}\rotatebox{90}{Type}  &\rotatebox{90}{Col. No}& Input Feature&\rotatebox{90}{Data Type}\\
% &No.& &  &  &No.&  &  \\
\hline
& 1 & duration & C & & 23 & Count & C\\
\cline{2-4} \cline{6-8}
& 2 & protocol\_type & S & & 24 & srv\_count & C\\
\cline{2-4} \cline{6-8}
& 3 & service & S & \multirow{8}{*}{\rotatebox{90}{Traffic}\rotatebox{90}{(Time Based)}} & 25 & serror\_rate & C\\
\cline{2-4} \cline{6-8}
\multirow{4}{*}{\rotatebox{90}{Basic}}& 4 & flag & S & & 26 & srv\_error\_rate & C\\
\cline{2-4} \cline{6-8}
& 5 & src\_bytes & C & & 27 & rerror\_rate & C\\
\cline{2-4} \cline{6-8}
& 6 & dst\_bytes & C & & 28 & srv\_rerror\_rate & C\\
\cline{2-4} \cline{6-8}
& 7 & land & S & & 29 & same\_srv\_rate & C\\
\cline{2-4} \cline{6-8}
& 8 & wrong\_fragment & C & & 30 & diff\_srv\_rate & C\\
\cline{2-4} \cline{6-8}
& 9 & urgent & C & & 31 & srv\_diff\_host\_rate & C\\
\cline{2-4} \cline{5-8}
& 10 & hot & C & & 32 & dst\_host\_count & C \\
\cline{1-4} \cline{6-8}
& 11 & num\_failed\_logins & C & & 33 & dst\_host\_srv\_count & C \\
\cline{2-4} \cline{6-8}
& 12 & logged\_in & S &\multirow{8}{*}{\rotatebox{90}{Traffic}\rotatebox{90}{(Host Based)}} & 34 & dst\_host\_same\_srv\_rate & C \\
\cline{2-4} \cline{6-8}
& 13 & num\_compromised & C & & 35 & dst\_host\_diff\_srv\_rate & C \\
\cline{2-4} \cline{6-8}
& 14 & root\_shell & C & & 36 & dst\_host\_same\_src\_port\_rate & C \\
\cline{2-4} \cline{6-8}
\multirow{4}{*}{\rotatebox{90}{Contents}}& 15 & su\_attempted & C & & 37 & dst\_host\_srv\_diff\_host\_rate & C \\
\cline{2-4} \cline{6-8}
& 16 & num\_root & C & & 38 & dst\_host\_serror\_rate & C \\
\cline{2-4} \cline{6-8}
& 17 & num\_file\_creations & C & & 39 & dst\_host\_srv\_serror\_rate & C \\
\cline{2-4} \cline{6-8}
& 18 & num\_shells & C & & 40 & dst\_host\_rerror\_rate & C \\
\cline{2-4} \cline{6-8}
& 19 & num\_access\_files & C & & 41 & dst\_host\_srv\_rerror\_rate & C \\
\cline{2-4} \cline{5-8}
& 20 & num\_outbound\_cmds & C \\
\cline{2-4}
& 21 & is\_hot\_login & S & \multicolumn{4}{c}{C means Continuous}\\ 
\cline{2-4}
& 22 & is\_guest\_login & S & \multicolumn{4}{c}{S means Symbolic}\\
\cline{1-4}
\multicolumn{4}{c}{}
\end{tabular}
}
\caption{Input features of the NSL-KDD dataset.}
\label{tab-inpf}
\end{table}

\subsection{Experimental setup}
\label{exp-setup}
The next step after the labels are generated is binarization. Detailed attention is needed to 
track the number of binary variables produced during 
this process.  In the case of {\it numeric} or {\it continuous} features, the number of binary variables generated is directly dependent on the number of cut-points. Thus,
if a feature is producing a large number of cut-points, it will increase the number of binary variables exponentially. For example,
if the number of cut-points is $100$, the total number of interval variables is $\binom{100}{2}=4950$ and after considering the level variables, the total number of binary variables created will be $4950+100=5050$. Consequently, the memory requirement will increase afterward to
an unmanageable level. On the other hand, a large number of cut-points indicate that the feature may not have much influence on the
classification of observations. Our strategy is to ignore such features  completely. Another set of features which
are having a fairly large number of cut-points are ignored partially. Given a feature $x$, if the number of cut-points is greater than or equal to $175$, we completely ignore the feature $x$ and if the number of cut-points is greater than or equal to $75$ but less than $175$, we ignore that feature partially by only generating the level variables. 
We have arrived at these thresholds after empirical analysis using the training data. 
List of features that have been fully or partially ignored are presented in the Table~\ref{tab-cutpt}.
{
\begin{table}[ht]
\center
\resizebox{1.0\columnwidth}{!}{
\begin{tabular}{|c|l|c|l|}
\hline
Col. Num. & Input feature & \#Cut-points & Ignored ?\\
\hline
1 & duration & 102 & Partially \\
\hline
5 & src\_bytes & 116 & Partially\\
\hline
23 & Count & 374 & {\bf Fully} \\
\hline
24 & srv\_count & 302 & {\bf Fully} \\
\hline
32 & dist\_host\_count & 254 & {\bf Fully} \\
\hline
33 & dist\_host\_srv\_count & 255 & {\bf Fully} \\
\hline
34 & dst\_host\_same\_srv\_rate & 100 & Partially \\
\hline
35 & dst\_host\_diff\_srv\_rate & 93 & Partially\\
\hline
36 & dst\_host\_same\_src\_port\_rate & 100 & Partially \\
\hline
38 & dst\_host\_serror\_rate & 98 & Partially\\
\hline
40 & dst\_host\_rerror\_rate & 100 & Partially\\ 
\hline
\multicolumn{4}{c}{}
\end{tabular}
}
\caption{Binarization: Ignored features of the NSL-KDD dataset.}
\label{tab-cutpt}
\end{table}
}

Another important aspect that we have incorporated into our design is the support of a pattern. Support of a positive (negative) pattern is $k$ if it covers $k$ positive (negative) observations and it should not cover any negative (positive) observation.  
Thus, the value of $k$ in Step~\ref{msup} of Algorithm~\ref{algo-pat} holds immense importance. In a previous implementation~\cite{bor00} the value of $k=1$
have been used, but it is observed during experiments that such a low support is generating a lot of patterns/rules having
little practical significance. Moreover, these patterns cause a lot of false positives during testing. An empirical
analysis helps us to fix the threshold at $k=100$. At this threshold, more than $95\%$ of the observations present in the training
dataset are covered by the generated patterns having degree up to $4$.  
\vspace*{-3mm}
\algrenewcommand\alglinenumber[1]{\tiny #1:}
\renewcommand{\thealgorithm}{1}
\floatname{algorithm}{Classifier}
\begin{algorithm}[ht]
	\centering
	\begin{algorithmic}[1]
		\tiny
		\STATE {{Input}: Observation $obs$ having $41$ features.}
		\STATE {{Output}: Class label $\mathcal L$.}
		\IF {$\neg(obs(36) \ge 0.0050) \land (obs(37) \ge 0.0050 \land obs(37) < 0.1150)$} %%% 1
		\STATE {$\mathcal L=1$} \COMMENT{$\mathcal L=1$ indicates normal behavior.}
		\ELSIF {$obs(5) \ge 28.50 \land \neg(obs(37) \ge 0.005 \land obs(37)< 0.915) \land strcmp(obs(r,3),'ftp\_data')$} %%% 2
		\STATE {$\mathcal L = 1$}
		\ELSIF  {$obs(5) \ge 28.5000 \land (obs(37) \ge 0.0050 \land obs(37)< 0.9150) \land obs(34) \ge 0.1950$} %%% 3
		\STATE {$\mathcal L=1$}
		\ELSIF {$obs(5) \ge 28.500000 \land \neg(obs(5) \ge 333.500000) \land obs(37) \ge 0.005000 \land obs(37) < 0.085000$} %%% 4
		\STATE {$\mathcal L=1$}   
		\ELSIF {$obs(5) \ge 28.500000 \land \neg(obs(34) \ge 0.645000) \land strcmp(obs(r,3),'ftp\_data')$} %%% 5  
		\STATE {$\mathcal L=1$}
		\ELSIF {$obs(6) \ge 0.500000 \land obs(37) \ge 0.005000 \land obs(37) < 0.915000 \land \neg(obs(40) \ge 0.005000)$}    %%% 6
		\STATE {$\mathcal L=1$}
		\ELSIF {$obs(6) \ge 0.500000 \land \neg(obs(5) \ge 333.500000) \land obs(5) \ge 181.500000$} %%% 7
		\STATE {$\mathcal L=1$}
		\ELSIF {$obs(5) \ge 333.500000 \land (obs(36) \ge 0.015000) \land obs(6) \ge 0.500000 \land obs(6) < 8303.000000$} %%% 8
		\STATE {$\mathcal L=1$}
		\ELSIF {$obs(5) \ge 333.500000 \land obs(34) \ge 0.645000 \land obs(6) \ge 0.500000 \land obs(6)< 8303.000000$} %%% 9
		\STATE {$\mathcal L=1$}
		\ELSIF {$\neg(obs(40) \ge 0.005000) \land obs(34) \ge 0.645000 \land \neg(obs(36) \ge 0.005000)$} %%$ 10
		\STATE {$\mathcal L=1$}
		\ELSIF {$\neg(obs(40) \ge 0.005000) \land obs(6) \ge 0.500000 \land obs(6) < 8303.000000 \land \neg(obs(34) \ge 0.045000)$} %%$ 11
		\STATE {$\mathcal L=1$}
		\ELSIF {$obs(5) \ge 28.50 \land obs(6) \ge 0.500 \land obs(36) \ge 0.015000 \land \neg(obs(10) \ge 0.500000 \land obs(10) < 29.000000)$} %%$ 12
		\STATE {$\mathcal L=1$}
		\ELSIF {$obs(5) \ge 28.50 \land obs(6) \ge 0.500 \land \neg(obs(40) \ge 0.005000) \land \neg(obs(10) \ge 0.500000 \land obs(10) < 29.000000)$} %%$ 13
		\STATE {$\mathcal L=1$}
		\ELSE
		\STATE {$\mathcal L=0$} \COMMENT{$\mathcal L=0$ indicates attack.}
		\ENDIF	
	\end{algorithmic}
	\caption{\small Details of SSIDS}
	\label{clsfier}
\end{algorithm}
\vspace*{-5mm}
\subsection{Experimental Results}
We have described all the steps required to design a classifier in the offline phase. Let us now summarize the outcome of the individual steps.
\newline
1.~{\it Labeling}: We have used $5000$ labeled observations for labeling  $20192$ unlabeled observations as described in Section~\ref{slad}. This step produces $22601$ labeled observations which have been used in the following steps to design the classifier.
\newline 
2.~{\it Binarize}: During this step, total $10306$ binary variables are produced and a binary dataset along with its class labels having size $22601 \times 10307$ is generated.  
\newline
3.~{\it Support Set Generation}: We have selected $21$ binary features according to their discriminating power.
\newline
4.~{\it Pattern Generation}: During pattern generation, we found $13$ positive and $7$ negative patterns.
\newline
5.~{\it Classifier Design}: We have developed a rule-based IDS using the $13$ positive patterns that are generated in the last step. 
Thus, the SSIDS contains $13$ rules. The details
of the SSIDS is available in the Classifier~\ref{clsfier}. The NSL-KDD dataset contains two test datasets: (i) KDDTest\textsuperscript{+} having $22,544$ observations, and (ii) KDDTest\textsuperscript{21} having $11,850$ observations. These two datasets are used to measure the accuracy of the
proposed SSIDS and the results related to the accuracy of the IDS is presented in Table~\ref{tab-res}. These results compare favorably with
the state of the art classifiers proposed in~\cite{rana17}, and the comparative results are presented in  Table~\ref{tab-comp}. It is evident that the proposed SSIDS outperforms the existing IDSs by a wide margin.
{
\begin{table}[ht]
\center
\resizebox{1.0\columnwidth}{!}{\bf
\begin{tabular}{|c|c|c|c|c|c|}
\hline
{\small Dataset} & {\small Accuracy} & {\small Precision} & {\small Sensitivity} & {\small F{1}-Score} & {\small Time in sec.}\\
\hline
 KDDTest\textsuperscript{+} & 90.91\% &0.9458 &0.8915 & 0.9179& 0.000156 \\
\hline
KDDTest\textsuperscript{21} & 83.92\% &0.9417 & 0.8564&0.8971& 0.000173\\
\hline
\multicolumn{3}{c}{}
\end{tabular}
}
\caption{Results related to KDDTest\textsuperscript{+} and KDDTest\textsuperscript{21}.}
\label{tab-res}
\end{table}
%\vspace*{-4mm}
}
\begin{table}[ht]
\center
\resizebox{0.80\columnwidth}{!}{
\begin{tabular}{|c|c|c|}
\hline
Classifiers\textsuperscript{\$} & \multicolumn{2}{c|}{Accuracy using Dataset(\%)}\\
\cline{2-3}
& KDDTest\textsuperscript{+} & KDDTest\textsuperscript{21} \\
\hline
 J48\textsuperscript{*} & 81.05 & 63.97 \\
\hline
Naive Bayes\textsuperscript{*} & 76.56 & 55.77 \\
\hline
NB Tree\textsuperscript{*} & 82.02 & 66.16 \\
\hline
Random forests\textsuperscript{*} & 80.67 & 63.25 \\
\hline
Random Tree\textsuperscript{*} & 81.59 & 58.51 \\
\hline
Multi-layer perceptron\textsuperscript{*} & 77.41 & 57.34 \\
\hline
SVM\textsuperscript{*} & 69.52 & 42.29 \\
\hline
Experiment-1 of ~\cite{rana17} & 82.41 & 67.06 \\
\hline
Experiment-2 of ~\cite{rana17} & 84.12 & 68.82 \\
\hline
LAD\textsuperscript{@} & {87.42} & {79.09}  \\
\hline
Proposed SSIDS & {\bf 90.91} & {\bf 83.92}  \\
\hline
\multicolumn{3}{l}{* ~Results as reported in~\cite{rana17}.}\\
\multicolumn{3}{l}{@ Classifier designed using dataset D\textsubscript{L} only by}\\
\multicolumn{3}{l}{~~~~omitting the `labeling' process.}\\
\multicolumn{3}{l}{\$~~All the classifiers use the same training}\\
\multicolumn{3}{l}{~~~~dataset, i.e., KDDTrain\_20percent.}\\
%\multicolumn{3}{l}{}
\end{tabular}
}
\caption{Performance Comparison between different Classifiers, IDSs and the proposed SSIDS.}
\label{tab-comp}
\end{table}
\vspace*{-10mm}
\section{Conclusion}
\label{sec-con}
The intrusion detection system (IDS) is a critical tool used to detect cyber attacks, and semi-supervised IDSs are gaining popularity
as it can enrich its knowledge-base from the unlabeled observations also. Discovering and understanding the usage patterns from the past observations
play a significant role in the detection of network intrusions by the IDSs. Normally, usage patterns establish a causal 
relationship among the observations and their class labels and the LAD is useful for such problems where we need to 
automatically generate useful patterns that can predict the class labels of future observations. Thus, LAD
is ideally suited to solve the design problems of IDSs. However, the dearth of labeled observations makes it 
difficult to use the LAD in the design of IDSs, particularly semi-supervised IDSs, as we need to consider the 
unlabeled examples along with the labeled
examples during the design of IDSs. In this effort, we have proposed a simple methodology to extend the classical LAD to 
consider unlabeled observations along with the labeled observations. We have employed the proposed technique successfully 
to design a new semi-supervised ``Intrusion Detection System'' which outperforms the existing semi-supervised IDSs by a wide margin both in terms of accuracy and detection time.               
\bibliographystyle{IEEEtranS}
\bibliography{semilad}
%
%\newpage
\section*{Appendix}
\vspace{-4mm}

\centering
\begin{table*}[h]
 \resizebox{1.8\columnwidth}{!}{
 \small
\thickmuskip = 0.25\thickmuskip
 \medmuskip = 0.25\medmuskip
 \thinmuskip = 0.25\thinmuskip
 \begin{tabular}{|c|c|c|c|c|c||c|c|c||c|c|c|c|c|c||c|}
 		  \hline
 			\rotatebox{90}{$A \geq 3.05 $} & \rotatebox{90}{$A \geq 2.45$} & \rotatebox{90}{$A \geq 1.65$} & \rotatebox{90}{$1.65 \leq A < 3.05 $} & \rotatebox{90}{$ 2.45 \leq A < 3.05 $} & \rotatebox{90}{$ 1.65 \leq A < 2.45 $ }&\rotatebox{90}{
 			$B\geq 2.95 $} & \rotatebox{90}{$B\geq 1.85$} & \rotatebox{90}{ $1.85 \leq B < 2.95 $} &
 			\rotatebox{90}{$C \geq 4.5 $} & \rotatebox{90}{$C \geq 3.3$} & \rotatebox{90}{$C \geq 1.9$} & \rotatebox{90}{$ 1.9 \leq C < 4.5 $} & \rotatebox{90}{$ 1.9 \leq C < 3.3 $} & \rotatebox{90}{$ 3.3 \leq C < 4.5 $} & \rotatebox{90}{Class} \\
 			\hline
 			$b_1$&$b_2$&$b_3$&$b_4$&$b_5$&$b_6$&$b_7$&$b_8$&$b_9$&$b_{10}$&$b_{11}$&$b_{12}$&$b_{13}$&$b_{14}$&$b_{15}$&$\mathcal L$\\
 			\hline
 			1&1&1&0&0&0&1&1&0&0&0&1&1&1&0&1\\
 			\hline
 			0&1&1&1&1&0& 0&0&0& 1&1&1&0&0&0 &1\\
 			\hline
 			0&0&0&0&0&0& 0&1&1& 0&1&1&1&0&1 &1\\
 			\hline
 			1&1&1&0&0&0& 0&0&0& 0&1&1&1&0&1 &0\\
 			\hline
 			0&0&1&1&0&1& 0&1&1& 0&0&0&0&0&0 &0\\
 			\hline
 			\multicolumn{16}{c}{}
 			\end{tabular}
 		 %\caption {Binary dataset generated from the Table~\ref{dset2} having $15$ binary variables from $b_1$ to $b_{15}$.}
 		%\label{bintab}
 }
 \caption{Binary dataset generated from the Table~\ref{dset2} having $15$ binary variables from $b_1$ to $b_{15}$.}
\label{bintab}
\end{table*}
\end{document}